%% file: lattice2011.tex
\documentclass{PoS}

\usepackage{epsfig,multicol}
\usepackage[final]{showkeys}
\usepackage{amsmath}
\usepackage{amssymb}
\input{mymacros}

\title{Towards an Effective Importance Sampling in Monte Carlo Simulations of a System with a Complex Action}

\ShortTitle{Effective Importance Sampling in Complex Action Systems}

\author{\speaker{Konstantinos N. Anagnostopoulos}
        \thanks{The work of K.N.A.\ was partially funded 
by the National Technical University of
Athens through the Basic Research Support Programmes 2009 and 2010.
The work of T.A. and J.N.\ is supported in part by Grant-in-Aid 
for Scientific Research 
(No.\ 23740211 for T.A. and 19340066, 20540286 for J.N.)
from Japan Society for the Promotion of Science.}\\
        Physics Department, 
        National Technical University of Athens, 
        Zografou Campus, 157--80 Zografou, Greece\\
        E-mail: \email{konstant@mail.ntua.gr}}

\author{Takehiro Azuma\\
        Institute for Fundamental Sciences, Setsunan University,
        17-8 Ikeda Nakamachi, Neyagawa, Osaka 572-8508, Japan\\
        E-mail: \email{azuma@mpg.setsunan.ac.jp}}

\author{Jun Nishimura\\
        High Energy Accelerator Research Organization (KEK) and
        Graduate University for Advanced Studies (SOKENDAI),  
        1-1 Oho, Tsukuba 305-0801, Japan\\
        E-mail: \email{jnishi@post.kek.jp}}

\abstract{
The sign problem is a notorious problem, which occurs in Monte Carlo
simulations of a system with a partition function whose integrand is
not positive. One way to simulate such a system is to use the
factorization method where one enforces sampling in the part of the
configuration space which gives important contribution to the
partition function. This is accomplished by using constraints on some
observables chosen appropriately and minimizing the free energy
associated with their joint distribution functions. These observables are
maximally correlated with the complex phase. Observables not in this
set essentially decouple from the phase and can be calculated without
the sign problem in the corresponding ``microcanonical''
ensemble. These ideas are applied on a simple matrix model with very
strong sign problem and the results are found to be consistent with
analytic calculations using the Gaussian Expansion Method.
}

\FullConference{The XXIX International Symposium on Lattice Field Theory - Lattice 2011\\
July 10-16, 2011\\
Squaw Valley, Lake Tahoe, California}

\begin{document}
\section{Introduction}

Monte Carlo simulation is an essential tool in the non perturbative
study of a quantum field theory and has provided a wealth of
information and insight in high energy physics. Its success lies in
the fact that algorithmic importance sampling of a very small fraction
of configurations gives a very accurate estimate of expectation values
of physical
observables. By stochastically constructing Markov chains of
configurations, the system quickly thermalizes and one is able to
sample the extremely small subset $\cal R$ of the configuration space
$\cal C$ that
gives the important contribution to the partition function. $\cal R$
is determined by the competition of entropy with the Boltzmann factor
of the action. For an
observable ${\cal O}$ whose value on the configuration $\mu$ is ${\cal
  O}[\mu]$, the Markov chain sample $\{\mu_1,\mu_2,\ldots,\mu_M\}$ is
constructed with sampling probability $P_\mu$ and $\vev{\cal O}$'s estimator is
\begin{equation}
\label{1.1}
{\bar{\cal O}} = 
\frac{\sum_{i=1}^M(P_{\mu_i})^{-1}{\cal O}[{\mu_i}] \ee^{-S[{\mu_i}]}}
     {\sum_{i=1}^M(P_{\mu_i})^{-1}                  \ee^{-S[{\mu_i}]}}\, .
\end{equation}
By choosing $P_\mu\propto \ee^{-S[{\mu_i}]}$, the above sum is
easily performed numerically without terms whose size varies {\it
  exponentially} with the system size\footnote{like e.g. in
  multihistogramming.}.  

This approach breaks down in the case where the action of the system
is complex and almost all configurations contribute a non positive
term in the partition function. The sample is constructed by
simulating a {\it phase quenched} 
model. One then mostly samples a subset ${\cal R}_0$ of $\cal C$ whose
overlap with $\cal R$ is exponentially small with system
size. The reason is that besides entropy and the Boltzmann factor of
the real part of the action, the fluctuations of the imaginary part of
the action
become a determining factor in the suppression of configurations. 
Moreover, if one takes a simple reweighting approach to
estimating $\vev{\cal O}$, one obtains
\begin{equation}
\label{1.2}
{\bar{\cal O}} = 
\frac{\sum_{i=1}^M{\cal O}[{\mu_i}] \ee^{i\Gamma[{\mu_i}]}}
     {\sum_{i=1}^M                  \ee^{i\Gamma[{\mu_i}]}}\, ,
\end{equation}
where $\Gamma[{\mu}]$ is the imaginary part of the value of the action
on $\mu$. Then, besides the small overlap, one has to sum terms that
oscillate wildly due to the exponential that appears in the numerator
and the denominator. The oscillations are much stronger away from the
stationary configurations of $\Gamma$ which is usually what happens when we
sample in ${\cal R}_0$.

An approach to studying such systems is the factorization method, originally
proposed in \cite{0108041}, used in \cite{Ambjorn:2003rr} and recently generalized in
\cite{multifac}. The key point is the selection of a maximal set of
observables 
\begin{equation}
\label{1.3}
\Sigma = \{{\cal O}_k | k=1,\ldots,n\} 
\end{equation}
which are strongly correlated with the phase $\ee^{i\Gamma}$. The idea
is that by solving saddle point equations for the minimum of the
``free energy'' associated with the distribution functions of these
observables, one can essentially determine the region $\cal R$. The
solutions are obtained by performing Monte Carlo simulations on
selected subspaces of $\cal C$ by constraining the values of these
observables.  Further calculations of any other observable ${\cal O}$
can be done by essentially sampling in $\cal R$ and their expectation
values can be estimated {\it without} the phase factor appearing in
\rf{1.2}. To be more specific, if $\{\bar x_1, \bar x_2, \ldots, \bar
x_n\}$ are the values\footnote{More precisely  $\bar x_k$  are equal
  to the values of ${\cal
    O}_k/\vev{{\cal O}_k}_0$ as it will be
  explained in the next paragraph.} of $\{{\cal O}_1,{\cal
  O}_2,\ldots,{\cal O}_n\}$ that are the solutions to the free energy
minimization equations, then one finds that
\begin{equation}
\label{1.4}
\vev{{\cal O}} \approx \vev{{\cal O}}_{\bar x_1, \bar x_2, \ldots, \bar x_n}
\end{equation}
where $\vev{\ldots}_{\bar x_1, \bar x_2, \ldots, \bar x_n}$ are
expectation values in a ``microcanonical'' system where
${\cal O}_k$ are constrained to be equal to $\bar x_k$. The right hand
side of \rf{1.4} {\it has no complex action} problem. The complex
action problem has been reduced to computing the solution $\{\bar
x_1,\ldots,\bar x_n\}$ and this is greatly improved by factorizing
the phase factor and taking advantage of its (hopefully) nice scaling
properties.

In this talk, the above general statements will be made concrete in a
specific example. A pedestrian's approach is adopted for presenting
the main ideas and the reader is referred to \cite{multifac} for the
details. A simple matrix model with very strong complex action
problem is studied and is shown how to compute the set $\Sigma$ and
the solution $\{\bar x_1,\ldots,\bar x_n\}$. Solving the equations
that give the stationary points of $\Gamma$ plays an important role in
determining $\Sigma$. The nice scaling properties of the distribution functions
are heavily used in the computation of the solution $\{\bar
x_1,\ldots,\bar x_n\}$.

Since the method can in principle be applied to any system with a
complex action problem, we hope that our exposition will contribute to
a successful application of the method to other interesting problems
in lattice field theory and elsewhere.

\section{The Model}

Matrix models have been studied extensively in
the context of non perturbative formulations of string theory and in
the study of gauge/gravity duality.
Monte Carlo simulations have contributed crucially in the study of the
large $N$ limit of supersymmetric matrix models \cite{Krauth:1998xh},
in providing first principle evidence of gauge/gravity duality and in
explaining the thermodynamics of certain black hole solutions in terms
of microscopic string degrees of freedom \cite{Hanada:2007ti}.

 The large $N$ limit of the IIB
matrix model \cite{Ishibashi:1996xs} has been proposed as a
non perturbative definition of string theory. In this model,
space-time emerges as the eigenvalue distribution of the bosonic
matrices which makes possible to study the scenario of {\it dynamical}
compactification of extra dimensions. This happens via spontaneous
symmetry breaking of the rotational symmetry (SSB) of the model.
Calculations using the Gaussian Expansion Method (GEM) support the
realization of such a scenario \cite{Nishimura:2001sx}. Monte Carlo
simulations can play an important role in confirming those results
from first principle calculations and in elucidating the mechanism that is
responsible for SSB.
 The strong fluctuations of the phase
$\ee^{i\Gamma}$ favour length scales over which spacetime extends
which are quite
different than in the phase quenched model 
\cite{0108041}. A simple matrix model that realizes the above scenario has
been proposed in \cite{0108070}. It is defined by the partition
function
\begin{equation}
\label{total_pf}
  Z = \int dA \, d \psi \, d {\bar \psi} \,
\ee^{-(S_{\rm b} + S_{\rm f})} \qquad\mbox{where}\qquad
  S_{\rm b} = \frac{1}{2} \, N \, \tr (A_{\mu})^{2} \, , \qquad 
  S_{\rm f} = - {\bar \psi}^{f}_{\alpha}\, (\Gamma_{\mu})_{\alpha \beta}\,
  A_{\mu} \psi^{f}_{\beta} \ .
\end{equation}
$A_{\mu}$ ($\mu= 1,2,3,4$) are $N \times N$ hermitian matrices, and
${\bar \psi}^{f}_{\alpha}$ and $\psi^{f}_{\alpha}$ ($\alpha=1,2$,
$f=1,\ldots,N_f$) are $N$-dimensional row and column vectors.  The
actions (\ref{total_pf}) have an $\textrm{SO}(4)$ symmetry, where the
bosonic variables $A_\mu$ transform as vectors and the fermionic
variables transform as Weyl spinors. Integrating out the fermions, we obtain
$Z = \int dA \, \eexp{-S_{\rm b}} \, Z_{\rm f}[A]$, 
where $Z_{\rm f}[A]= (\det {\cal D})^{N_{\rm f}}$ and ${\cal D} =
\Gamma_{\mu} A_{\mu}$ is a $2N \times 2N$ matrix. The fermion
determinant $\det {\cal D}$ for a single flavor is complex in general
but it turns out to be real for configurations with $A_4=0$ and that
the phase of the determinant {\it becomes stationary} for
configurations with $A_4=A_{3}=0$ \cite{Nishimura:2000ds}.
We take the large-$N$ limit with $r=N_{\rm f}/N$ fixed, which
corresponds to the Veneziano limit.

The order parameters of SSB of $\textrm{SO}(4)$ are the expectation
values of the eigenvalues of the ``moment of inertia tensor''
$ T_{\mu \nu} = \frac{1}{N} \tr (A_{\mu} A_{\nu})\, .$
These are ordered as  $\lambda_{1} > \lambda_{2} >  \lambda_{3} >
\lambda_{4} > 0$ and if their VEVs turn out to be unequal in the large-$N$ limit, it
signals the SSB of $\textrm{SO}(4)$. GEM calculations \cite{0412194}
show that $\textrm{SO}(4)$ breaks down to $\textrm{SO}(2)$ for all
$r>0$ and Monte Carlo simulations are consistent with this result \cite{multifac}.

 The ``phase quenched model'' is defined by
\beq
  \label{eq:z0}
  Z_0 = \int dA \, e^{-S_{0}[A]} \, , \qquad 
  S_{0}[A] = 
S_{\rm b}[A] - N_{\rm f} \log |\det {\cal D}[A]|\, .
\eeq
and in this case SSB is absent. In fact $\vev{\lambda_n}_0 =
1+\frac{r}{2}$ for all $n=1,2,3 , 4$ 
where the VEVs $\vev{\ \cdot \ }_0$ are taken with respect to
$(\ref{eq:z0})$. In the following, we will study the eigenvalues of
$T_{\mu \nu}$ normalized to their phase quenched expectation values
and we denote $\tilde\lambda_n=\lambda_n/\vev{\lambda_n}_0$.
Deviation of $\vev{\tilde\lambda_n}$ from 1 indicates a strong effect of the
complex phase.
In order to simulate  (\ref{total_pf}) we rewrite it as
$ Z = \int dA \, \eexp{-S_{0}[A]} \, \eexp{i \Gamma[A]}$,
where due to nonzero $\Gamma[A]$ for generic configurations, the
system turns out to have a very strong complex action problem. Due to
the stationarity of the phase for lower dimensional configurations,
the eigenvalues $\tilde\lambda_n$ are chosen for the application of
the factorization method. 

We study the distribution function
$\rho(x_1,x_2,x_3,x_4) = \vev{\prod_{k=1}^4 \delta(x_k-\tl_k)}$
and the corresponding one in the phase quenched model 
$\rho^{(0)}(x_1,x_2,x_3,x_4) = \vev{\prod_{k=1}^4  \delta(x_k-\tl_k)}_0$. 
By defining the ``microcanonical ensemble''
$Z_{x_1,x_2,x_3,x_4}=\int\, dA \ee^{-S_0}\prod_{k=1}^4 \delta(x_k-\tl_k)\, ,$
we define the function $w(x_1,x_2,x_3,x_4)=\vev{\ee^{i\Gamma}}_{x_1,x_2,x_3,x_4}$
and one finds that
\begin{equation}
\label{2.2}
\rho(x_1,x_2,x_3,x_4)
=\frac{1}{C}\rho^{(0)}(x_1,x_2,x_3,x_4)w(x_1,x_2,x_3,x_4)\, ,
\end{equation}
where $C=\vev{\ee^{i\Gamma}}_0$. The minimum of ${\cal
  F}(x_1,x_2,x_3,x_4)=-\ln \rho(x_1,x_2,x_3,x_4)$ is the estimator for
$\{\vev{\tl_1},\vev{\tl_2},\vev{\tl_3},\vev{\tl_4}\}$. One has to solve the saddle point equations
\begin{equation}
\label{2.3}
\frac{1}{N^2}f_n^{(0)}(x_1 , x_2 , x_3 , x_4) = - 
 \frac{\partial}{\partial x_n} \Phi (x_1 , x_2 , x_3 , x_4)
 \, , \quad n=1,2,3,4
\end{equation}
where $f_n^{(0)}(x_1 , x_2 , x_3 , x_4)=\frac{\partial}{\partial x_n}\ln \rho^{(0)}(x_1 , x_2 , x_3 , x_4)$,
$\Phi(x_1 , x_2 , x_3 , x_4) = \lim_{N\to\infty}\frac{1}{N^2} \ln w(x_1 , x_2 , x_3 , x_4)$.
The nice scaling properties of $f_n^{(0)}(x_1 , x_2 , x_3 , x_4)$ and $\Phi(x_1 , x_2 , x_3 , x_4)$ allow for
extrapolations in $N$ and $x_n$. It is also important to note that the
error in $\vev{\tl_n}$ does not propagate exponentially in $N$.

\begin{figure}
    \epsfig{file=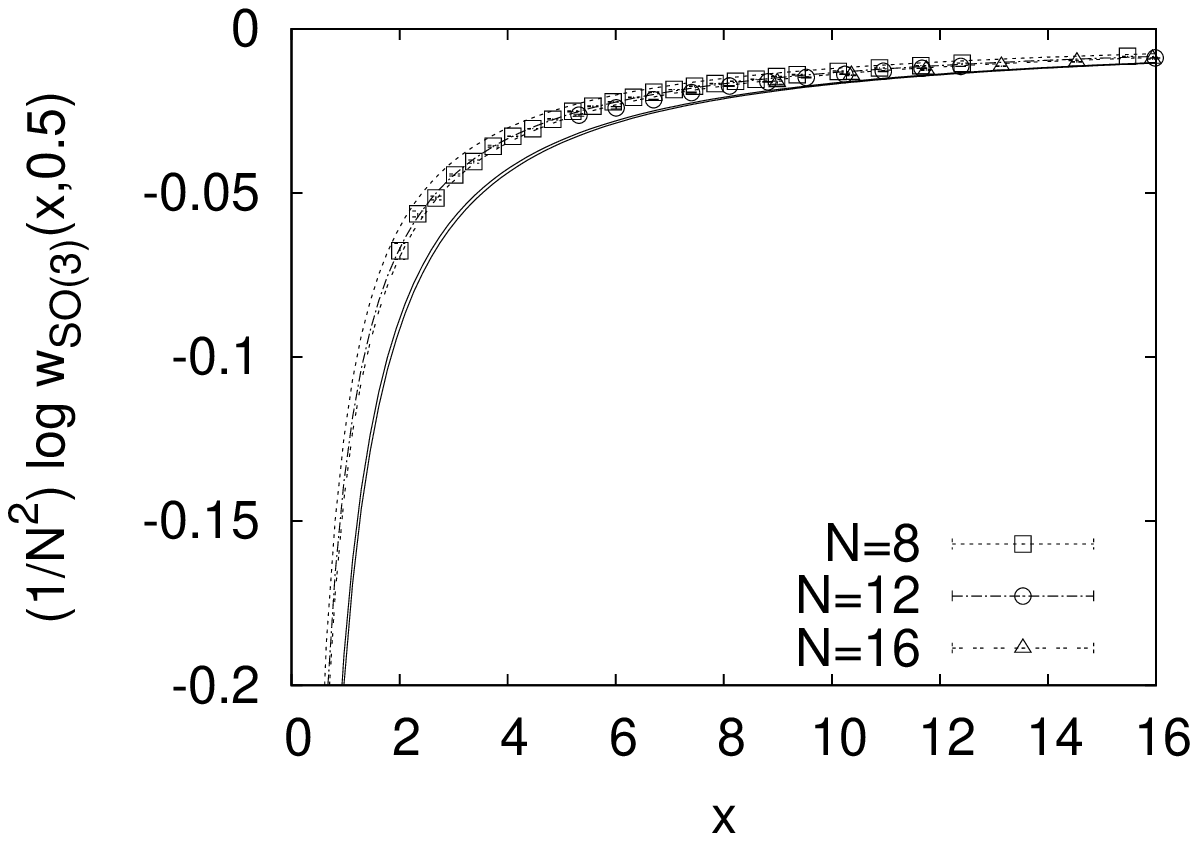,width=7.4cm}
    \epsfig{file=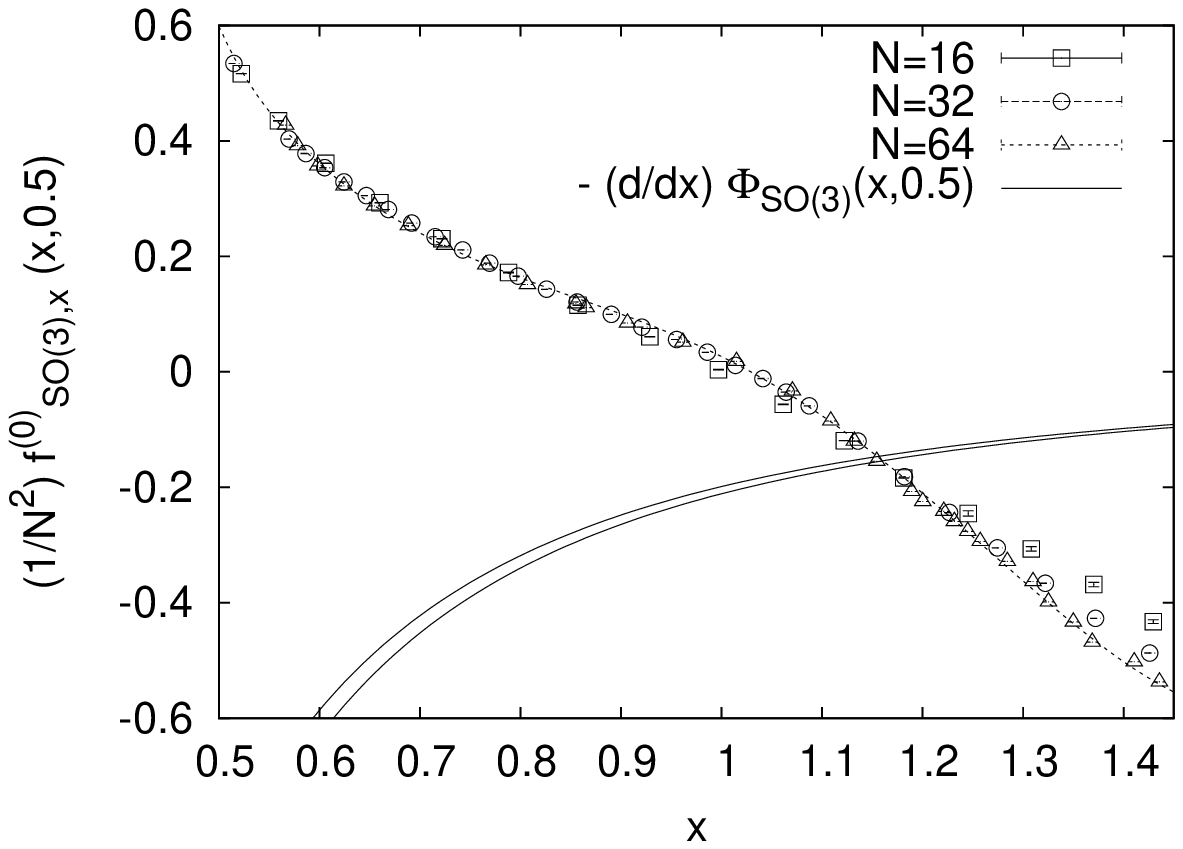,width=7.4cm}
    \caption{
(Left) The function
$\frac{1}{N^2} \log w_{\textrm{SO(3)}} (x,0.5)$ 
is plotted 
against $x$
for $N=8,12,16$.
The solid lines represent
the function
$\Phi_{\textrm{SO(3)}} (x,0.5)$
obtained by extrapolation to
$N=\infty$ as described in \protect\cite{multifac}. 
(Right) The function
$\frac{1}{N^2} f^{(0)}_{\textrm{SO(3)},x} (x,0.5)$ 
is plotted 
against $x$ 
for $N=16,32,64$.
The solid lines represent
$- \frac{\partial}{\partial x} \Phi_{\textrm{SO(3)}} (x,0.5)$
obtained from the plot on the left.}
   \label{no-phase-li}
\end{figure}

Looking for the full solution of \rf{2.3} using Monte Carlo is a
formidable task. Therefore, guided by the GEM results in
\cite{0412194} we look for SO(3) and SO(2) symmetric solutions. For
lack of space we discuss the SO(3) symmetric vacuum and the interested
reader is referred to \cite{multifac} for the SO(2) case. We take
 $x_1=x_2= x_3 > 1> x_4$ and  define the reduced functions 
$
\rho^{(0)}_{\textrm{SO(3)}} (x,y) = \rho^{(0)} (x,x,x,y)
$,
$
w_{\textrm{SO(3)}} (x,y) = w(x,x,x,y)
$,
$
f_{{\rm SO(3)},x_i}^{(0)}(x,y)=\frac{\partial}{\partial  x_i}\ln\rho^{(0)}_{\textrm{SO(3)}}(x,y)$ and $
\Phi_{\rm SO(3)}(x,y) = \lim_{N\to\infty}\frac{1}{N^2}\ln  w_{\rm SO(3)}(x,y)
$. 
Equations \rf{2.3} become
\begin{equation}
\label{2.4}
\frac{1}{N^2}\, f_{{\rm SO(3)},x}^{(0)} (x, y) = - \frac{\partial}{\partial x} \Phi_{\rm SO(3)} (x, y)
\, ,\qquad
\frac{1}{N^2}\, f_{{\rm SO(3)},y}^{(0)} (x, y) = -
\frac{\partial}{\partial y} \Phi_{\rm SO(3)} (x, y)\, .
\end{equation}
In the following we fix $r=N_f/N=1$ and from the GEM results
\cite{0412194}  we expect that the SO(3)
symmetric solution is 
$(x_{{}_{\rm GEM}},y_{{}_{\rm GEM}})\approx (1.17,0.50)$. We first fix
$y=y_{{}_{\rm GEM}}=0.50$ and solve for $x$. The results can be seen
in fig.~\ref{no-phase-li}. We find $x=1.152(3)$ which is consistent
with $x_{{}_{\rm GEM}}=1.17$.
\begin{table}
\begin{center}
\begin{tabular}{|c|c|c|c|c|}
\hline
ansatz & \multicolumn{2}{|c|}{SO(3)} &\multicolumn{2}{|c|}{SO(2)}\\
\hline
method &  factorization  & GEM &  factorization & GEM \\ 
\hline
\vev{\tilde{\lambda}_1} & ---      & 1.17  & ---      & 1.4\\
\vev{\tilde{\lambda}_2} & ---      & 1.17  & 1.373(2) & 1.4\\
\vev{\tilde{\lambda}_3} & 1.151(2) & 1.17  & 0.649(4) & 0.7\\
\vev{\tilde{\lambda}_4} &  0.59(2) & 0.5   & 0.551(2) & 0.5\\
\hline
\end{tabular}  
\end{center}
\caption{\label{tab:lam} The results for the normalized eigenvalues 
$\vev{\tilde{\lambda}_n}$ for $r=1$
obtained by the factorization method
for the SO(3) and SO(2) symmetric vacua.
The dash implies that the result should be the same as the 
one below in the same column due to the imposed symmetry.
We also show the GEM results obtained at $N=\infty$
in ref.~\cite{0412194}.}
\end{table}
Next we fix $x=x_{{}_{\rm GEM}}=1.17$ and solve for $y$. We repeat the
same procedure for the SO(2) symmetric vacuum and our results are
summarized in Table \ref{tab:lam}.

Including more variables in the analysis is straightforward. 
If $(X,X,Y,Z)$ is the absolute maximum of $\rho(x_1,x_2,x_3,x_4)$, consider the 
``microcanonical ensemble'' $\vev{\cdot}_{X,X,Y,Z}$ and for an operator ${\cal O}$ 
define
$\rho_{\cal O}(x) = 
\frac{
\langle \delta ( x - \widetilde{\cal O}) \, \ee^{i \Gamma}
\rangle_{X , X , Y, Z}
}{
\langle \ee^{i \Gamma}
\rangle_{X , X, Y , Z}
}$,
$\rho_{\cal O}^{(0)}(x) = 
\langle \delta ( x - \widetilde{\cal O})  
\rangle_{X , X , Y, Z}
$
where $\widetilde{\cal O}={\cal O}/\vev{{\cal O}}_0$.
Then
$\rho_{\cal O}(x)  = \frac{1}{C'} \,
\rho_{\cal O}^{(0)}(x)  \, w_{\cal O} (x)$
where $C'= \langle \ee^{i \Gamma} \rangle_{X,X,Y,Z}$ and
$
w_{\cal O} (x) = 
\frac{
\langle \delta ( x - \widetilde{\cal O}) \, \ee^{i \Gamma}
 \rangle_{X , X , Y, Z}
}{
\langle \delta ( x - \widetilde{\cal O})
\rangle_{X , X, Y , Z}
} 
$.
Next we determine $\vev{\cal O}$ from
$\frac{1}{N^2} f^{(0)}_{\cal O} = -\frac{d}{d x} \Phi_{\cal O}(x)\, ,$
where $f^{(0)}_{\cal O}=\frac{d}{d x} \ln \rho^{(0)}_{\cal O}(x)$ and
$\Phi_{\cal O}(x) = \lim_{N\to\infty} \frac{1}{N^2}\ln w_{\cal O}(x)$.
If $\vev{\cal O}$ does not shift much from $\vev{{\cal O}}_{X,X,Y,Z}$ then the overlap
problem for $\cal O$ is not severe. Then one can show that \cite{multifac}
\begin{equation}
\label{2.6}
\vev{{\cal O}}_{X,X,Y,Z}\approx
\frac{\vev{{\cal O}\ee^{i\Gamma}}_{X,X,Y,Z}}{\vev{\ee^{i\Gamma}}_{X,X,Y,Z}}\, .
\end{equation}
i.e. correlation of $\cal O$ and $\ee^{i\Gamma}$ 
within $\vev{\cdot}_{X,X,Y,Z}$ is small. The advantage of this
relation is that the phase factors out and 
one can calculate the expectation value of $\cal O$
within the microcanonical ensemble without the sign problem.
\begin{figure}
    \epsfig{file=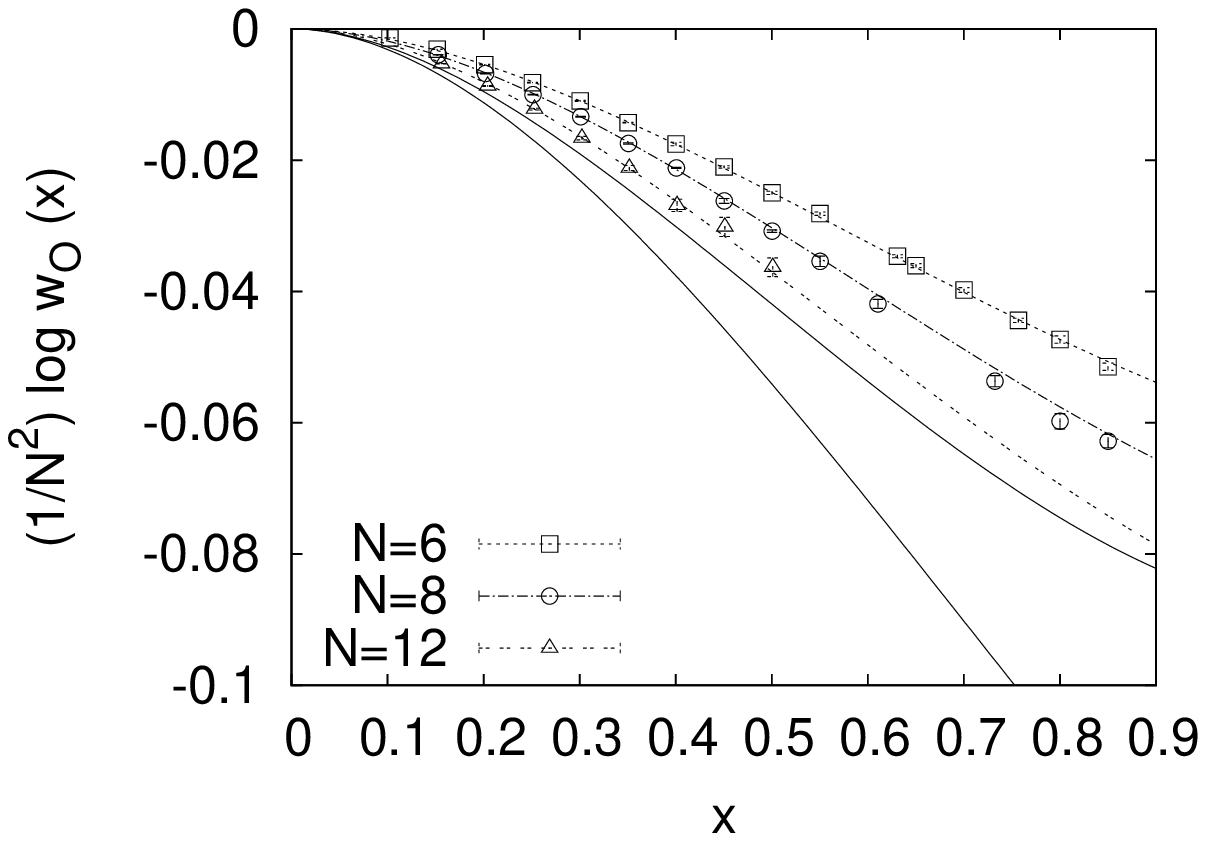,width=7.4cm}
    \epsfig{file=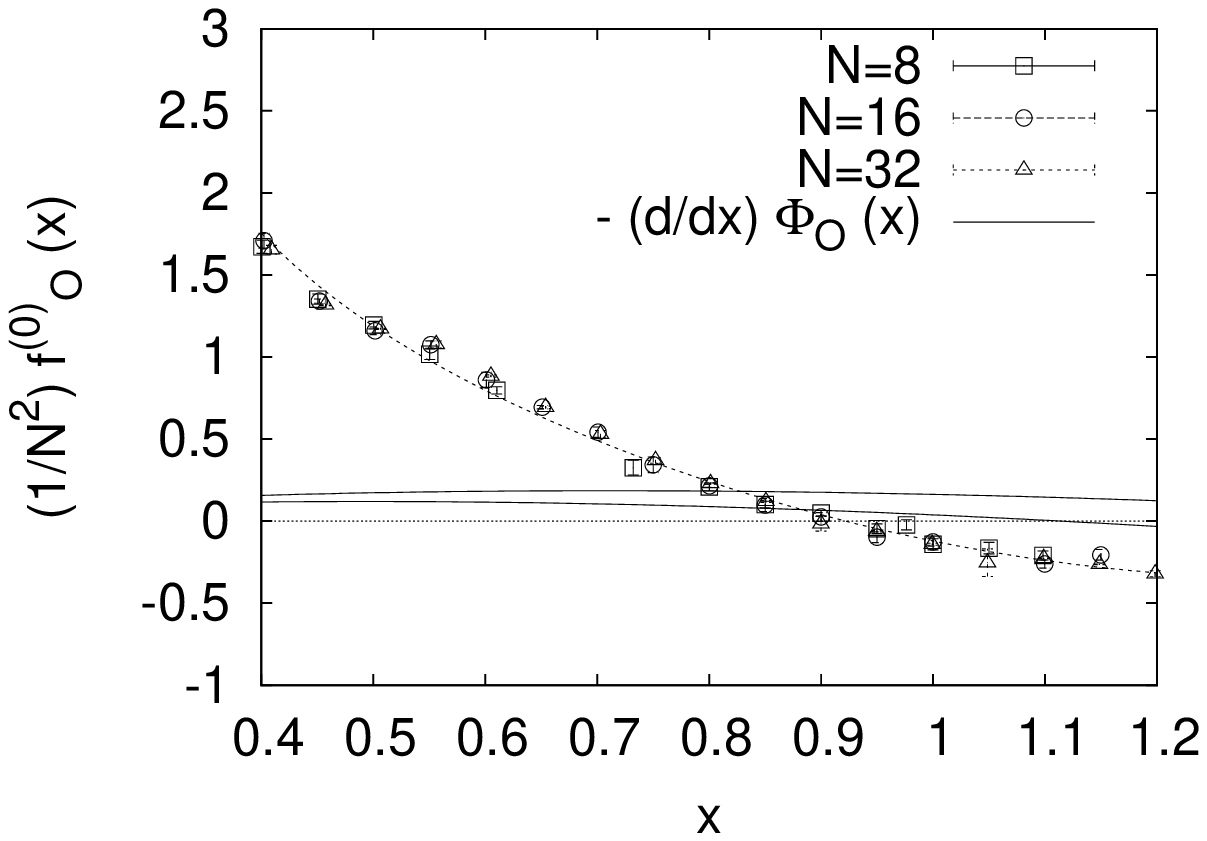,width=7.4cm}
    \caption{(Left) The function
$\frac{1}{N^2}  \log w_{{\cal O}}(x)$ is plotted against $x$
for $N=6,8,12$.
We also plot the function
$\Phi_{{\cal O}}(x)$ 
obtained by extrapolation to
$N=\infty$ as described in \protect\cite{multifac}.
The two solid lines represent the margin of error.
(Right) The function 
$\frac{1}{N^2} \frac{d}{dx} \log \rho^{(0)}_{{\cal O}}(x)$ 
is plotted for $N=8,16,32$. 
We also plot $- \frac{d}{dx} \Phi_{{\cal O}}(x) $
obtained from the plot on the left.
}
   \label{t:trF2}
\end{figure}

We test the above relations by considering the observable 
 ${\cal O}= - \frac{1}{N} \sum_{\mu \neq \nu}\tr [A_\mu , A_\nu]^2 $. As
$x\ll 1$ the dominant configurations are approximately simultaneously
diagonalizable $[A_\mu , A_\nu]\approx 0$ and $\det{\cal D}\ge
0$. Therefore $\cal O$ can potentially have strong correlations with
the phase factor.  From fig.~\ref{t:trF2} (Left) 
we find that
$\frac{1}{N^2} \log w_{\cal O}(x)$ approaches zero 
for $x\rightarrow 0$ as expected. From fig.~\ref{t:trF2} (Right) 
we find that the effect of the phase 
is to shift the estimate of 
$\langle \widetilde{\cal O} \rangle$
by $\Delta x = 0.07(3)$.
On the other hand, the standard deviation of 
the distribution $\rho^{(0)}_{\cal O}(x)$ is estimated
as $\sigma \sim 0.7/N$ from the slope of the function
plotted in fig.~\ref{t:trF2} (Right) around $x\sim 0.92$.
This means that the deviation $\Delta x$ is 
$\lesssim 2 \, \sigma$ for $N \le 16$.
Thus, the remaining overlap problem 
associated with this observable is practically small.
This is consistent with the fact that we were able to reproduce
the GEM result by constraining only the four observables 
$\lambda_n$ ($n=1,2,3,4$). 

\section{Conclusions}

In this work, we have applied the factorization method to the Monte
Carlo study of a matrix model with strong complex action problem. It
has been extended to include more than one observables in order to
eliminate the overlap problem and sample effectively the theory's
configuration space. A maximal set $\Sigma$ of such observables has
been determined that has significant correlations with the complex
phase. By constraining sampling to configurations that lie in the
neighbourhood of the solution to the minimization of the free energy
of the distribution functions of the observables in $\Sigma$, all
other observables can be computed without the complex action
problem. The complex action problem has been reduced to solving the
saddle point equations, which is difficult, but not impossible if one
takes advantage of the nice scaling properties of the factors in the
distribution functions. Solving the equations for the stationary
configurations of $\Gamma$ played an important role in determining
$\Sigma$.

Our results turn out to be consistent with analytical calculations
using the GEM and confirm SSB of rotational invariance in the
distribution of eigenvalues of bosonic matrices. This is related to
the problem of dynamical compactification of extra dimensions in
string theory which will be studied further in the IIB matrix model as
well as in simpler matrix models with supersymmetry.

The steps followed in this study are considered to be generic and we
hope that a similar approach can be applied to many interesting
systems.


\end{document}

%% file: mymacros.tex
\newcommand{\bel}{\begin{equation}\label}

\newcommand {\beq}{\begin{equation}}
\newcommand {\eeq}{\end{equation}}
\newcommand {\beqa}{\begin{eqnarray}}
\newcommand {\eeqa}{\end{eqnarray}}
\newcommand {\bc}{\begin{center}}
\newcommand {\ec}{\end{center}}
\newcommand {\tr}{{\rm tr\,}}

\newcommand {\ee}{\mbox{e}}

\newcommand {\vev}  [1]{\ensuremath{\langle #1 \rangle}}
\newcommand {\rf}   [1]{(\ref{#1})}
\newcommand {\eexp} [1]{\ensuremath{\mbox{e}^{#1}}}
\newcommand {\tl}      {{\tilde\lambda}}

\def\vs5{\vspace*{5mm}}
\def\vs1{\vspace*{1cm}}
\def\vs2{\vspace*{2cm}}
\def\hs5{\vspace*{5mm}}
\def\hs1{\hspace*{1cm}}
\def\hs2{\hspace*{2cm}}
\def\vs50{\vspace*{50mm}}
\def\vs20{\vspace*{20mm}}

\def\tr{\hbox{tr}}

\newcommand\jhep [3]{{{\it J. High Energy Phys.\ }{\bf #1} (#2) #3}}

\newcommand\ptp  [3]{{{\it Prog.\ Theor.\ Phys.\ }{\bf #1} (#2) #3}}

